\documentclass[twocolumn]{aastex62}

\newcommand{\feh}{$\mbox{[Fe/H]}$}

\received{June, 2020}
\revised{--}
\accepted{--}

\submitjournal{ApJL}

\shorttitle{Southern Hemisphere SkyMapper Metallicity Map}
\shortauthors{Chiti, Mardini, Frebel, \& Daniel}

\newcommand{\teff}{\ensuremath{T_\mathrm{eff}}}
\newcommand{\logg}{\ensuremath{\log\,g}}

\begin{document}

\title{The Metal-Poor Metallicity Distribution of the Ancient Milky Way}

\correspondingauthor{Anirudh Chiti}
\email{achiti@mit.edu}

\author[0000-0002-7155-679X]{Anirudh Chiti}
\affil{Department of Physics and Kavli Institute for Astrophysics and Space Research, Massachusetts Institute of Technology, Cambridge, MA 02139, USA}

\author[0000-0001-9178-3992]{Mohammad K.\ Mardini}
\affiliation{Key Lab of Optical Astronomy, National Astronomical Observatories, Chinese Academy of Sciences, Beijing 100101, China}
\affiliation{Institute of Space Sciences, Shandong University, Weihai 264209, China}

\author[0000-0002-2139-7145]{Anna Frebel}
\affiliation{Department of Physics and Kavli Institute for Astrophysics and Space Research, Massachusetts Institute of Technology, Cambridge, MA 02139, USA}

\author{Tatsuya Daniel}
\affiliation{Department of Physics, Brown University, Providence, RI 02912, USA}

\begin{abstract}

We present a low metallicity map of the Milky Way consisting of $\sim$111,000 giants with $-3.5 \lesssim$ [Fe/H] $\lesssim -$0.75, based on public photometry from the second data release of the SkyMapper survey.
These stars extend out to $\sim$7\,kpc from the solar neighborhood and cover the main Galactic stellar populations, including the thick disk and the inner halo.
Notably, this map can reliably differentiate metallicities down to [Fe/H] $\sim -3.0$, and thus provides an unprecedented view into the ancient, metal-poor Milky Way. 
Among the more metal-rich stars in our sample ([Fe/H] $> -2.0$), we recover a clear spatial dependence of decreasing mean metallicity as a function of scale height that maps onto the thick disk component of the Milky Way. 
When only considering the very metal-poor stars in our sample ([Fe/H] $< -$2), we recover no such spatial dependence in their mean metallicity out to a scale height of $|Z|\sim7$ kpc.
We find that the metallicity distribution function (MDF) of the most metal-poor stars in our sample ($-3.0 <$ [Fe/H] $< -2.3$) is well fit with an exponential profile with a slope of $\Delta\log(N)/\Delta$[Fe/H] = 1.52$\pm$0.05, and  shifts to $\Delta\log(N)/\Delta$[Fe/H] = 1.53$\pm$0.10 after accounting for target selection effects. 
For [Fe/H] $< -2.3$, the MDF is largely insensitive to scale height $|Z|$ out to $\sim5$\,kpc, showing that very and extremely metal-poor stars are in every galactic component. 

\end{abstract}



\section{Introduction} 
\label{sec:intro}
 
Many studies have focused on comprehensively characterizing the nature of the Milky Way's galactic components, which include the bulge \citep{hca+15}, the thin and thick disk \citep{dfa+18}, the metal-weak tail of the thick disk, and the inner and outer halo \citep{cbl+07} through determining spatial properties (e.g., scale heights, densities), stellar chemical abundances (e.g., metallicities), kinematic properties, and when possible, dating using age measurements. The halo is of particular interest for galactic archaeology studies that e.g., aim to understand the galaxy's early evolution.  
Firstly, a relatively large portion of the halo is composed of ancient very and extremely metal-poor stars ($\mbox{[Fe/H]}<-2.0$ and $<-3.0$, respectively), which are pivotal for studies of early chemical evolution \citep{fn+15}.
Secondly, theoretical simulations of galaxy formation \citep{bk+05} have shown that the halo bears the signatures of the Milky Way's assembly from smaller ``building block" galaxies.

Astrometric data from the \textit{Gaia} mission \citep{gaia+16, gaia+20} has shown e.g., the existence of stellar kinematic signatures as a result from earlier accretion events \citep{bee+18, mvi+19, zmb+20} but there remains the need for precise metallicities \textit{down to the extremely metal-poor regime} ([Fe/H] $\sim -3.0$) for many stars for a full chemodynamic picture of early structures still present in the current Milky Way. Recent spectroscopic and photometric surveys have began to remedy this. Spectroscopic surveys for which metallicities have been derived for large samples of stars include LAMOST \citep{zzc+12}, RAVE \citep{sgm+20}, GALAH \citep{dfb+15},  SDSS/SEGUE \citep{ewa+11, yrn+09}, and APOGEE \citep{msf+17}. Photometric surveys are far more complete than spectroscopic surveys and thus offer better prospects for mapping chemical structure because the relative rarity of extremely metal-poor stars means that these are generally poorly represented in any sample. \citet{isj+08} created a metallicity map of the Northern sky based on SDSS broadband photometry of $\sim5.6$ million stars. The disk and halo as well as the Monoceros stream cleanly differentiate in metallicity vs. height above the Galactic plane, confirming the galactic components and stellar populations. However, the SDSS photometry only enabled metallicity determination down to $\mbox{[Fe/H]}\sim-2.0$.

More recent photometric surveys, such as the SkyMapper Southern Sky Survey \citep{ksb+07} and the Pristine survey \citep{smy+17}, have instead developed the use of metallicity-sensitive filters surrounding the Ca II K line that enable metallicity measurements in the very and even extremely metal-poor regime.
These have led to notable results, including identification of the most metal-poor stars in the halo \citep{sab+18, dbm+19} and the bulge \citep{hca+15, asm+20}, and the metallicity distribution function of extremely metal-poor stars \citep{ksm+20}.

Additional spatial maps of the metallicity of the Northern hemisphere were presented in \citep{ab+20}, analogous to those in \citet{isj+08}, but extending to lower metallicities ([Fe/H] $\sim -3.0$).
Similarly, spatial maps of the Southern sky as a function of metallicity were presented in \citet{cwm+19} and \citet{hcy+19} using photometry from Data Release 1.1 from the SkyMapper Survey \citep{wol+18}.
However, these maps only claimed metallicity precision down to [Fe/H] $\sim-2.0$ or [Fe/H] $\sim -2.5$, and were limited by the relatively shallow magnitudes in SkyMapper DR1.1. 

Here we explore the oldest and most metal-poor stellar population spread across the galaxy with a series of metallicity maps of the Southern sky based on the recently released SkyMapper Southern Sky Survey Data Release 2 (DR2) \citep{owb+19}. 
The metallicity sensitive imaging filters \citep{bbs+11} enable a significant metallicity resolution down to $\mbox{[Fe/H]}\sim -3.3$. 
Using techniques developed in \citet{cfj+20} to extract precise metallicities from the intermediate-band SkyMapper $v$ and broadband $u,g,i$ filters, we derived photometric metallicities for every giant star with $g < 17$ and $\mbox{[Fe/H]}<-0.75$ in the full catalog (Chiti et al. ApJS subm.). We here focus on a subset of well-selected $\sim111,000$ metal-poor red giants with quality metallicity measurements to map out the ancient components of the Galaxy.

\section{Observational Data and Photometric Metallicities} 
\label{sec:obs}

The SkyMapper Southern Sky Survey DR2 contains $\sim500$ million astrophysical sources \citep{owb+19}. 
We downloaded photometry for every star brighter than $g = 17$ in this catalog, de-reddened the data following the bandpass coefficients on the SkyMapper website\footnote{http://skymapper.anu.edu.au/filter-transformations/}, and derived photometric stellar parameters of these stars (log\,g, [Fe/H]) following the methods presented in \citet{cfj+20}.
In summary, this involved matching the observed photometry to a grid of synthetic photometry spanning a broad range of stellar parameters (1.0 $<$ log\,$g < 3.0$, 4000\,K $<$ \teff\ $<$ 5700\,K, $-4.0 < $\feh\ $< -0.75$). 
We excluded all stars with initial\footnote{We note that $\sim5\%$ of stars have photometric metallicities up to 0.1\,dex higher than [Fe/H] $= -0.75$ due to a metallicity correction based on the spatial location of stars (see Section~2.3 of Chiti et al. ApJS subm.).} photometric [Fe/H] $\geq -0.75$, [Fe/H] $\leq -3.75$, and {\logg} $< 3.0$, and $0.35 < g-i < 1.2$ (corresponding to our {\teff} limits) to largely limit our sample to metal-poor red giant stars. 
We furthermore excluded stars with random uncertainties of $>$0.5\,dex in their photometric metallicity and those in regions of high reddening (galactic latitude $|b| < 10^{\circ}$ and E(B$-$V) $>$ 0.35 in \citealt{sfd+98}).
This resulted in a sample of 593,668 stars.
Finally, we find that we systematically overestimate the metallicities of very metal-poor ([Fe/H] $< -2.0$) stars when $g-i < 0.65$, upon comparison to high-resolution samples \citep{jkf+15, erf+20}.
To avoid this issue, we conservatively only retain stars with $g-i > 0.65$, resulting in 308,702 stars.
We refer the reader to Chiti et al. (ApJS, subm.) for further details on the full catalog.

We adopted photogeometric distances to every star in our sample from \citet{Bailer-Jones21} over using inverted parallax-based distances (see \citealt{Mardini_2019a} for a recent discussion). 
To ensure distance measurements of sufficient quality, we only included stars that have distance uncertainties less than the 20\% level.
We then excluded stars with absolute SkyMapper $g$ magnitude $> 5$ and a location on a color-magnitude diagram consistent with a metallicity [Fe/H] $> -0.75$ (Chiti et al. ApJS Subm.), which further limits contamination from cool dwarf stars and more metal-rich interlopers.
This resulted in a final sample of 111,149 stars extending down to $g\sim17$.

To validate the metallicities in our sample, we compared our metallicities to those from multiple other surveys. 
Details are given in Chiti et al. (ApJS subm.), but we list relevant comparisons here.
For $-2.5 < $ {\feh} $< -0.75$, when only considering stars with uncertainties on their photometric metallicities $< 0.5$\,dex, our metallicities are on average lower than those reported in the LAMOST survey by 0.05\,dex. 
The standard deviation of the residuals between our photometric metallicities and those in the LAMOST survey is $0.25$\,dex, indicating great agreement. 
Similarly, upon comparing to metallicities from the combined APOGEE and GALAH surveys over the same metallicity regime, we find, on average, lower metallicities by 0.21\,dex. 
The metallicity residuals have a standard deviation of 0.25\,dex.

\begin{figure}[t!]
\includegraphics[width =\columnwidth]{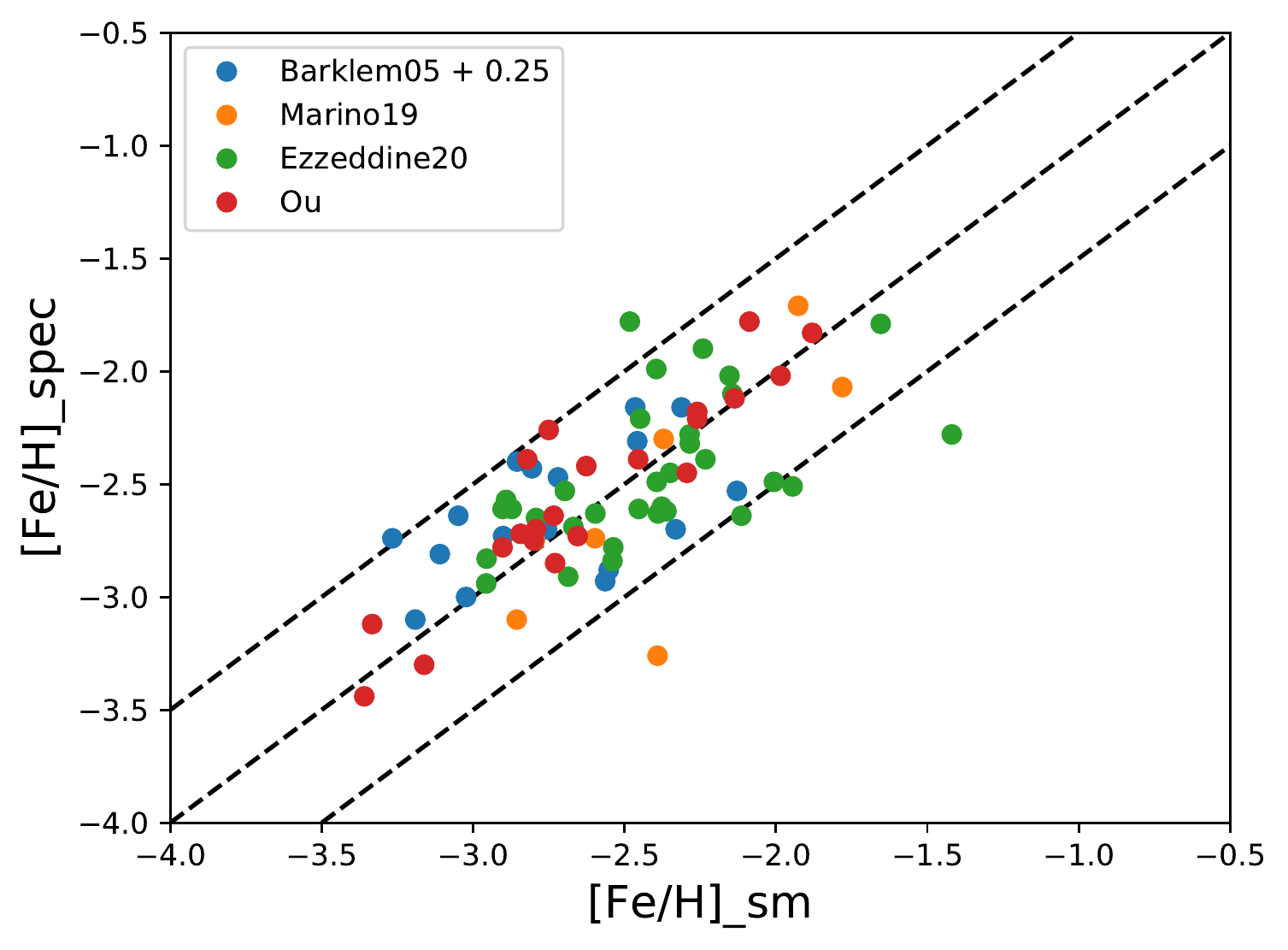}
\caption{Comparison of [Fe/H] metallicities between our photometric SM results and high-resolution measurements of stars in \citet{bcb+05, mda+19, erf+20} and X. Ou et al. (in prep).
The agreement is excellent, with $\Delta$[Fe/H] = 0.04$\pm$0.03. The standard deviation of the residuals between the metallicities is 0.29\,dex. 
Dashed lines indicate $\pm0.5$\,dex to guide the eye, and the metallicities in \citet{bcb+05} have been increased by 0.25\,dex to account for an empirical correction on {\teff} \citep{fcj+13}.}
\label{fig:hires}
\end{figure}

We demonstrate that our metallicities are robust when [Fe/H] $< -2.5$ following high-resolution observations of 74 initial metal-poor candidates selected from SkyMapper DR1.1 and DR2 data \citep{wol+18, owb+19} as an early test sample.
We derived their stellar parameters following standard spectral analysis techniques \citep{fcj+13} for these stars based on Magellan/MIKE \citep{bsg+03} high-resolution ($R \sim 22,000$) snapshot spectra that were reduced using the CarPy pipeline \citep{k+03}. 
Full results will be reported in X. Ou et al. (in prep) but we show results (red points) in Figure~\ref{fig:hires} for the 21 stars that we recovered in the DR2 catalog after our quality cuts. 
The comparison is excellent, with an average metallicity offset between the two samples of 0.09\,dex and a standard deviation in the residuals of 0.17\,dex. 
In Figure~\ref{fig:hires}, we also compare our values to the high-resolution samples of to \citet{bcb+05}, \citet{mda+19}, and \citet{erf+20} for stars with covering the range of our photometric metallicities ([Fe/H] $>$ $-3.75$).
Results largely follow what is presented in the comparisons to APOGEE and LAMOST and X. Ou et al. (in prep), but the \citet{bcb+05} comparison shows an offset of 0.25\,dex, largely due to them lacking an empirical correction on {\teff} \citep{fcj+13}.
The standard deviation of the residuals of our photometric metallicities and the high-resolution spectroscopic studies is 0.29\,dex.

We note that in Chiti et al. ApJS subm., we find that the completeness of our sample is independent of metallicity down to [Fe/H] $\sim -3.0$. 
Consequently, when presenting quantitative analysis (e.g., fitting an exponential slope) of the metallicity distribution function of our sample, we only include photometric metallicities down to [Fe/H] $= -3.0$ in the fitting procedures.

An additional bias is related to target selection effects, since metal-poor stars are brighter than metal-rich stars at the same effective temperature. 
Consequently, our sample may be preferentially biased toward low-metallicity stars at larger distances as our quality cuts will exclude more metal-rich stars at larger distances. 
As an illustration of this effect, stars right at the threshold of our distance quality cut (uncertainties at the 18\% to 20\% level) have a median distance of 6.4\,kpc when $-1.25 <$ [Fe/H] $< -0.75$, but a median distance of 8.9\,kpc when $-2.75 <$ [Fe/H] $<-2.25$.
However, this selection preference toward metal-poor stars is not significant within a distance of $\sim$5.0\,kpc, as we find that at distances between 4.0\,kpc and 5.0\,kpc, a roughly constant fraction of $<1$\% of stars are excluded from our full catalog by our quality cut as a function of metallicity. 
A weak trend with metallicity appears at distances between 5.0\,kpc and 6.0\,kpc, in which 6\% of stars are excluded with higher metallicities ([Fe/H] $> -1.25$), but only 2\% of stars are excluded with low metallicities ([Fe/H] $< -1.75$). 
This suggests a slight preferential selection of low metallicity stars at distances beyond $\sim$5.0\,kpc, but no bias at distances nearer.


\section{Metal-Poor Stars in the Milky Way}

\begin{figure}[t!]
\includegraphics[width =\columnwidth]{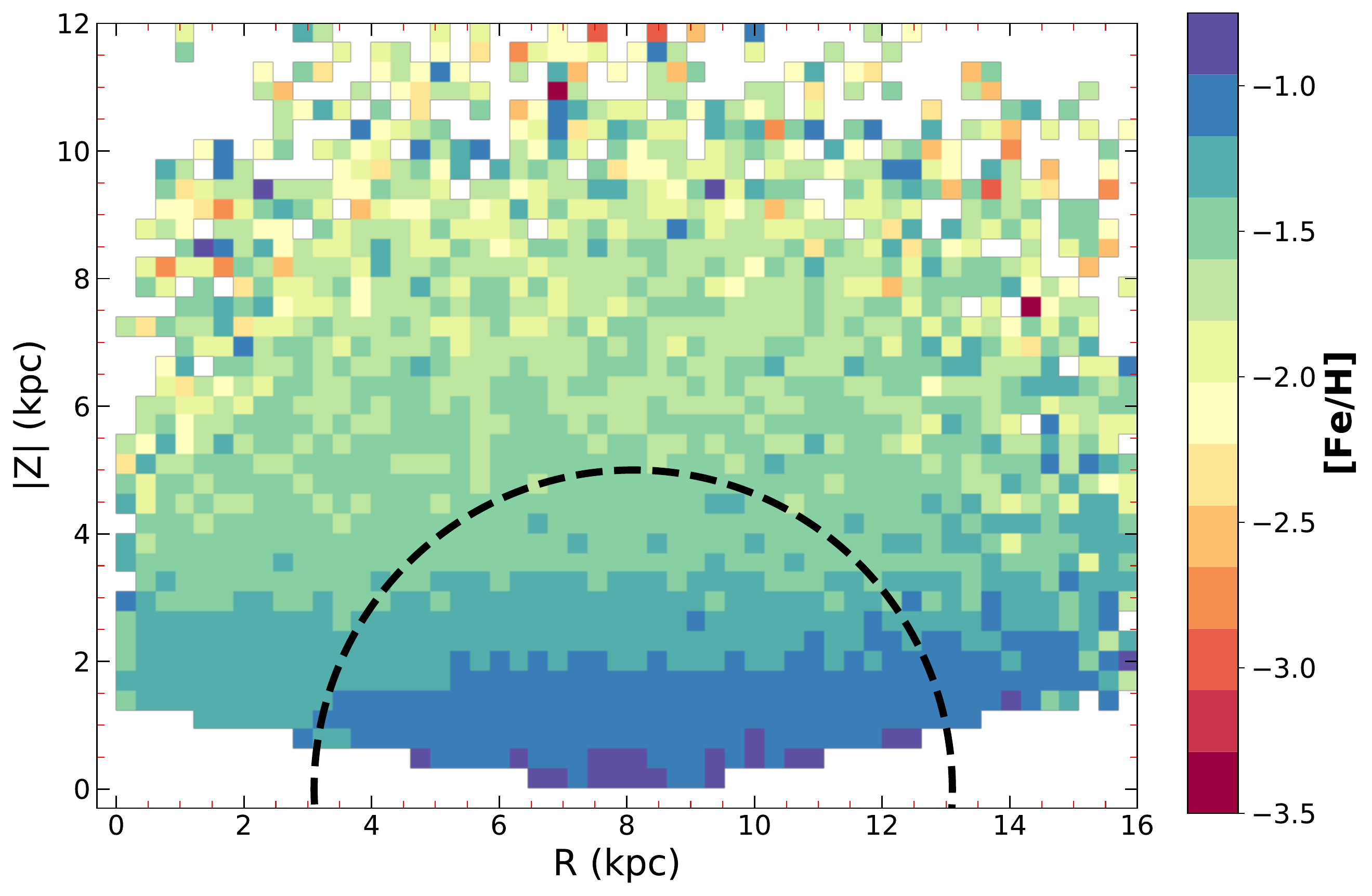}
\caption{Top: Low metallicity map of stars with $\mbox{[Fe/H]}\leq-0.75$ in the R-$|Z|$ plane, where R is the galactocentric distance and $|Z|$ is the absolute value of the height above the galactic plane. 
A clear, decreasing metallicitiy gradient is apparent as a function of $|Z|$, and the Milky Way disk is visible via a drop-off in average metallicity above $|Z| \sim 2.8$\,kpc. 
Each cell in the plot has dimensions of 300\,pc by 300\,pc, and is colored by the mean metallicity of stars contained within its region.
The dashed semicircle includes stars with distances $<$ 5.0\,kpc, within which target selection effects do not appreciably bias the metallicity of the sample (see Section~\ref{sec:obs} for discussion).}
\label{fig:metallicity_map}
\end{figure}

Our low-metallicity map ($\mbox{[Fe/H]}\leq-0.75$) of the Milky Way is shown in Figure~\ref{fig:metallicity_map} as a function of scale height ($|Z|$) and distance from the Galactic center ($R$). 
Using \texttt{astropy} \citep{astropy}, we transformed the coordinates of each star into the $R$, $Z$ plane, assuming the Sun's location of $8.1$\,kpc away from the Galactic center and 22\,pc above the Galactic midplane \citep{Gravity_Collaboration}. 
Each location of this map is colored by the average metallicity of stars in the corresponding spatial bin (300\,pc by 300\,pc in size). 
The disk/halo separation is immediately easy to recognize, due to the transition of the average metallicity from $\mbox{[Fe/H]}\gtrsim-1.25$ to $\mbox{[Fe/H]}\lesssim-1.25$ above $|Z|\sim2.8$\,kpc. 
This compares well to the metallicity map presented in \citet{isj+08}, whom also find a sharp break at $|Z| \sim3$\,kpc above which the mean metallicity drops below $\mbox{[Fe/H]}\sim -1.25$. 

\begin{figure*}[t!]
\includegraphics[width =\textwidth]{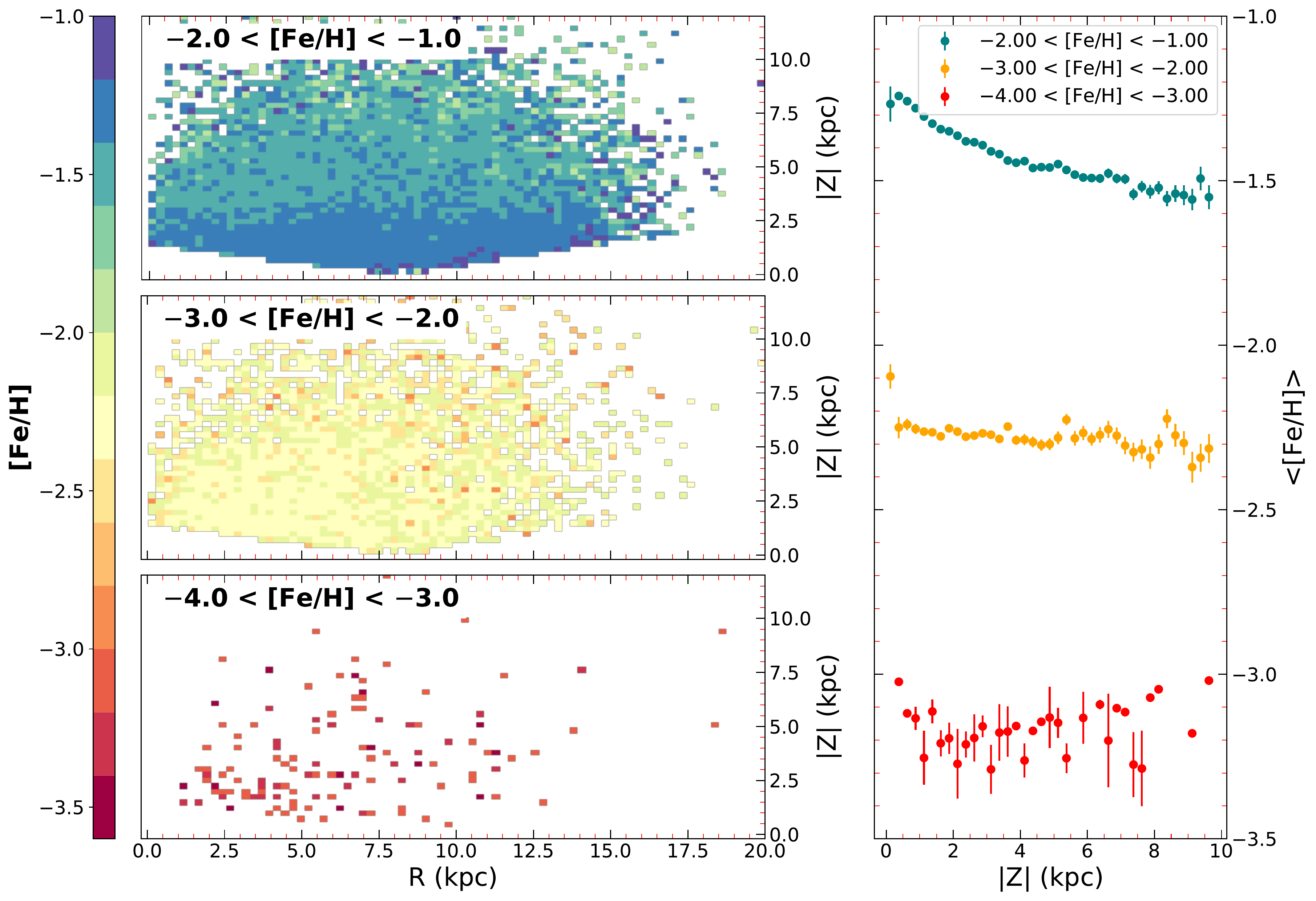}
\caption{Left panels: Low metallicity maps of stars in different metallicity ranges. The thick disk is easily apparent among stars with $-2.0 < \mbox{[Fe/H]}< -1.0$. No significant structure is present among stars with $\mbox{[Fe/H]}< -2.0$.
Right panel: Mean metallicities of stars as a function of scale height $|Z|$, within 0.25\,kpc bins in $|Z|$ that correspond to the respective left panels. The disk region induces a clear metallicity gradient (top) that makes way for no significant change in [Fe/H] with increasing $|Z|$.}
\label{fig:metallicity_bins}
\end{figure*}

In Figure~\ref{fig:metallicity_bins}, we show metallicity maps for several metallicity ranges (1\,dex increments from $\mbox{[Fe/H]}=-1.0$ to $\mbox{[Fe/H]}=-4.0$) to display the structure of the Galaxy at progressively lower metallicities. 
The Milky Way disk is still clearly noticeable in our highest metallicity bin ($-2.0 < \mbox{[Fe/H]} < -1.0$), along with a diffuse population of more metal-poor stars ($-2.0 <\mbox{[Fe/H]} < -1.5$) that follow no obvious spatial structure. 
Continuing this trend, the stars in the two lowest metallicity bins ($-3.0 < \mbox{[Fe/H]} < -2.0$ and  $-4.0 < \mbox{[Fe/H]} < -3.0$) appear to show no spatial correlation with their mean metallicity.

To illustrate this trend, in the right panel of Figure~\ref{fig:metallicity_bins}, we plot the mean metallicity of each of these metallicity bins as a function of scale height $|Z|$.
While the mean metallicity of our most metal-rich bin shows a decreasing mean metallicity as a function of $|Z|$, the more metal-poor bins do not, suggesting no significant spatial correlation with metallicity below $\mbox{[Fe/H]}=-2.0$. 
Overall, this figure confirms that the most metal-poor stars are spatially distributed in all of the Galactic components, but their search in certain components (e.g., the disk) is simply impeded by the wealth of the more metal-rich stars.
In fact, 0.8\% of stars in our sample have $\mbox{[Fe/H]}< -2.5$ when $|Z| < 3$\,kpc, but comprise 2.7\% of stars when $|Z| > 3$\,kpc.

\begin{figure*}[htbp!]
\includegraphics[width =18cm]{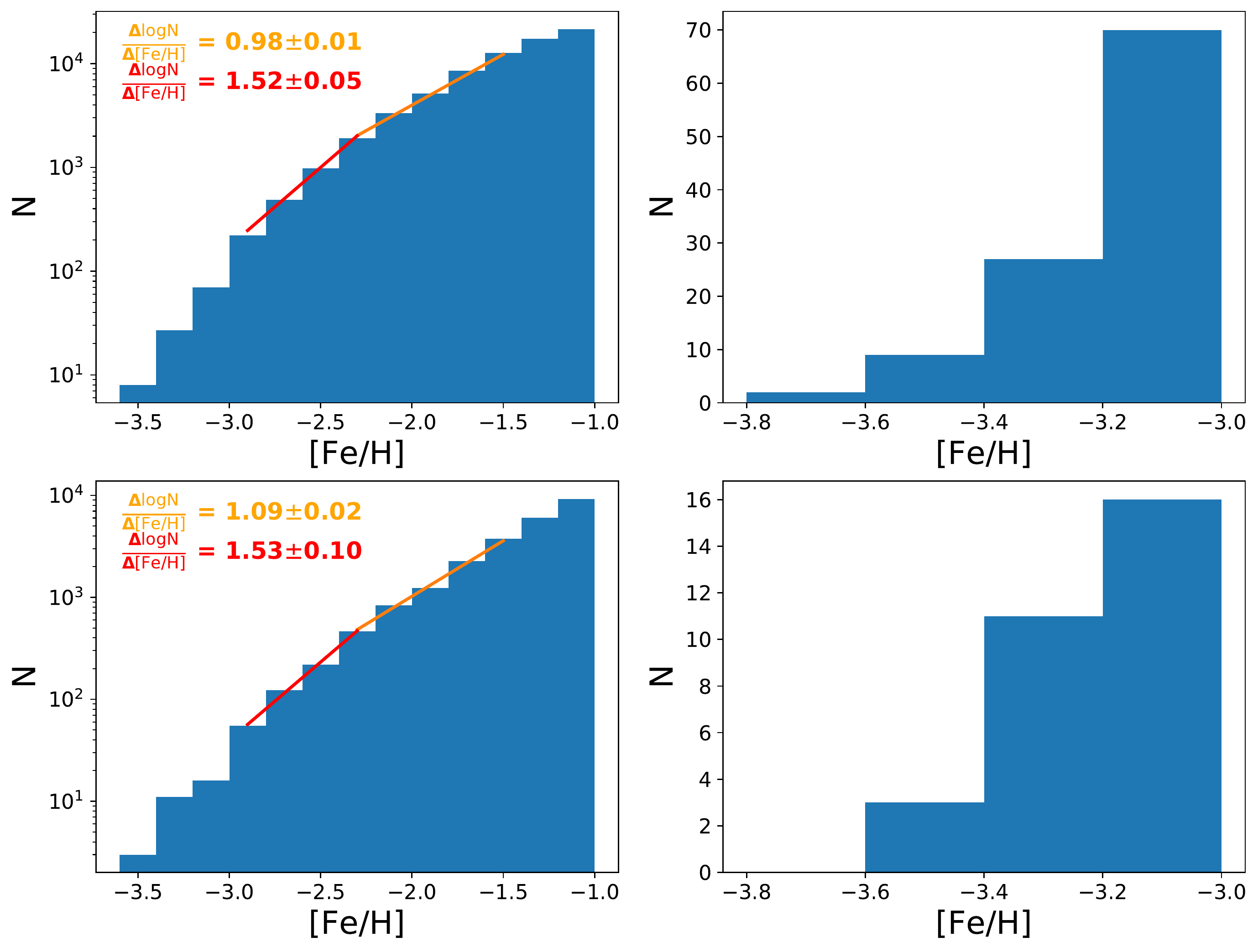}
\caption{Top left: Metallicity distribution function (MDF) below $\mbox{[Fe/H]} < -0.75$ of $\sim$122,000 metal-poor giants. Best-fitting exponential fits are shown, $(\Delta\text(\log N))/(\Delta\text{[Fe/H]})$ = 0.98 $\pm 0.01$ when $-2.3 <$ [Fe/H] $ < -$1.5 and $(\Delta\text(\log N))/(\Delta\text{[Fe/H]})$ = 1.52 $\pm$ 0.05 when $-3.0<$ [Fe/H] $< -2.3$.
Fitting for the more metal-poor regime only includes stars  $\mbox{[Fe/H]} > -3.0$, to avoid incompleteness effects (see Section~\ref{sec:obs} for discussion). 
Top right: MDF of stars with [Fe/H] $<-3.0$ in our sample.
Bottom left and right: Same as top row, but only including stars with distances $<$5.0\,kpc to minimize target selection effects (see Section~\ref{sec:obs} for discussion).}
\label{fig:mdf}
\end{figure*}

In Figure~\ref{fig:mdf}, we present the metallicity distribution function (MDF) of our sample of metal-poor giants. 
We now include stars with a loosened distance quality cut (uncertainty of $<40\%$), since those giants with unreliable distances are most likely to be associated with the outer halo and populate the most metal-poor regime of the MDF.
As mentioned in Section~\ref{sec:obs}, we only include stars with [Fe/H] $> -3.0$ when fitting profiles to the MDF, given incompleteness effects when [Fe/H] $< -3.0$.

Below $\mbox{[Fe/H]}=-$1.5, the MDF smoothly decreases exponentially towards lower metallicities.
Down to $\mbox{[Fe/H]}\sim-2.3$, this behavior is well described by an exponential function with a slope of $(\Delta\text(\log N))/(\Delta\text{[Fe/H]})$ = 0.98 $\pm$ 0.01.
This slope was derived through a MCMC procedure by implementing the likelihood function for a truncated exponential distribution (equation 8.1.17 in \citealt{cohen1991}) in the python package emcee \citep{emcee}.
Then, the derived exponential slope was converted from $\Delta\text(\ln N)/\Delta\text{[Fe/H]}$, as was returned from this likelihood function, to $\Delta\text(\log N)/\Delta\text{[Fe/H]}$.
This slope suggests that the number of the metal-poor stars drops by a factor of $\sim$10 for every dex decrease in [Fe/H], supporting conventional wisdom that a factor of $\sim10$ fewer stars exist for each 1\,dex drop in metallicity.
Below $\mbox{[Fe/H]}=-2.3$, the slope steepens to $(\Delta\text(\log N))/(\Delta\text{[Fe/H]})$ = 1.52 $\pm$ 0.05, implying that stars with metallicities below this cutoff are progressively more difficult to find as this corresponds to a drop by a factor of $\sim$33 for every dex decrease in [Fe/H].
This translates into a frequency of 1.5\% of finding stars with $\mbox{[Fe/H]}\sim-3.0$ and 0.2\% when $\mbox{[Fe/H]}\sim-3.5$ among stars with [Fe/H] $< -1.5$ within our sample.
Overall, the MDF is remarkably smooth down to the lowest [Fe/H], as was also found in previous studies based on smaller samples \citep{fcn+06, schoerck09, ynb+13, dbm+19}, and other large photometric studies \citep{abs+15, ksm+20}. 

Only including stars with distances $<$ 5.0\,kpc, to account for target selection effects, results in a slope of $(\Delta\text(\log N))/(\Delta\text{[Fe/H]}))$ = 1.09 $\pm 0.02$ when $-$2.3 $<$ [Fe/H] $<$ $-$1.5 and $(\Delta\text(\log N))/(\Delta\text{[Fe/H]}))$ = 1.53 $\pm$ 0.10 when $-3.0 <$ [Fe/H] $< -2.3$. 
We note that the slope at the more metal-poor end is negligibly different, suggesting a robustness to target selection effects.
However, the slope at the more metal-rich end is notably steeper. 
While this steepening plausibly does arise from addressing target selection effects that exclude metal-rich stars at large distances, it may also be caused by the exclusion of spatially distant stars that have an underlying MDF that is preferentially metal-poor \citep[e.g.,][]{isj+08}.

Our derived exponential slope of the MDF is consistent with that of previous spectroscopic studies \citep{schoerck09, afs+14}, but is steeper than what is found in others \citep{abs+15, ksm+20}.
By taking the relative ratio of the MDF in \citet{schoerck09}  at [Fe/H] = $-3.0$ and [Fe/H] = $-2.3$ suggests an exponential slope of $(\Delta\text(\log N))/(\Delta\text{[Fe/H]}) \sim 1.73$, which is comparable to our result of $\Delta\log(N)/\Delta$[Fe/H] = 1.52$\pm$0.05 or 1.53$\pm$0.10 in that regime.
A similar calculation for the MDF presented in \citet{afs+14} suggests an exponential slope of $(\Delta\text(\log N))/(\Delta\text{[Fe/H]}) \sim 1.66$.
However, the same analysis returns an exponential slope of 0.98 from \citet{abs+15}, and \citet{ksm+20} present a value of 1.0 $\pm$ 0.1, with which our derived slope is somewhat discrepant.
This suggests that underlying selection effects between our samples might be affecting the slope. 
For instance, this study only focuses on relatively cool giant stars within $\sim7$\,kpc, whereas other studies purely focus on stars in the halo ($>6$\,kpc away).

\begin{figure*}[htbp!]
\includegraphics[width =\textwidth]{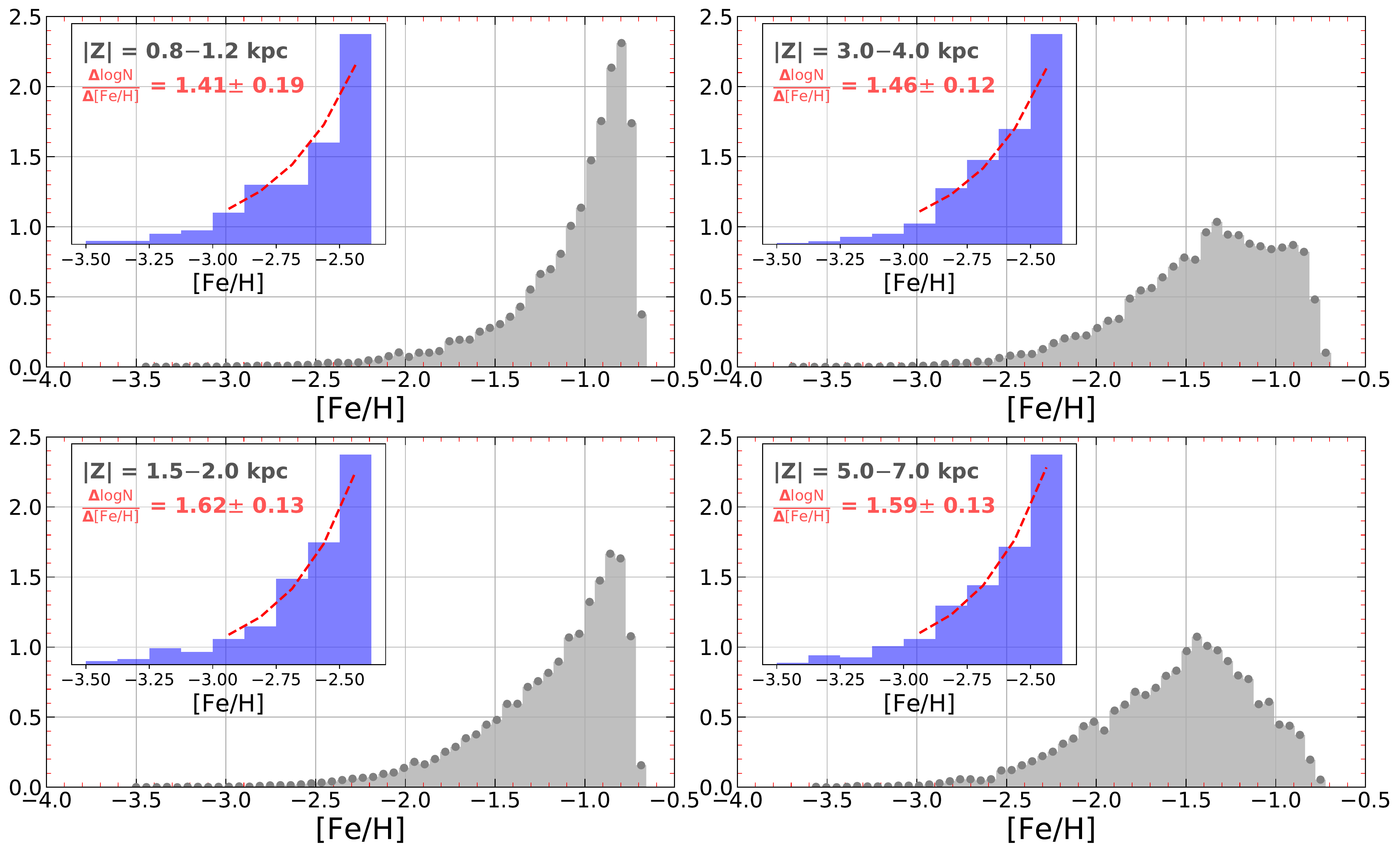}
\caption{Metallicity distribution of $\sim$122,000 stars, arranged as a function of distance from the Galactic plane in four different ranges, mimicking the inner thick disk (top left), the outer thick disk (bottom left), the metal-weak tail of the thick disk (top right), and the halo (bottom right).
Insets show the MDFs of stars with [Fe/H] $< -2.3$, with the best fitting exponential slope indicated.}
\label{fig:metallicity_distribution}
\end{figure*}

We also investigate how the shape of the MDF varies as a function of height above the galactic plane. 
In Figure~\ref{fig:metallicity_distribution}, we show MDFs for four different scale heights. 
The closest range is $0.8<|Z|<1.2$\,kpc, followed by $1.5<|Z|<2.0$\,kpc, $3.0<|Z|<4.0$\,kpc, and $5<|Z|<7$\,kpc.
At low $|Z|$, the thick disk component clearly dominates the higher [Fe/H] end of the distribution, but wanes beyond $2$\,kpc.
By $3$\,kpc the more metal-rich population associated with the thick disk population no longer dominates the distribution, and the metal weak tail of the thick disk becomes apparent around $\mbox{[Fe/H]}\sim-1.3$.  
Finally, beyond 5\,kpc, the halo dominates, with the MDF peaking at $\mbox{[Fe/H]}\sim-$1.5 and showing a significant metal-poor tail. 
The more metal-rich thick disk population appears to be entirely removed.

In Figure~\ref{fig:metallicity_distribution}, we also fit the very metal-poor tail of the MDF ([Fe/H] $< -2.3$) with an exponential slope to see if variations occur with scale height. 
We find four slopes: 
$(\Delta\text(\log N))/(\Delta\text{[Fe/H]})$ = 1.41 $\pm$ 0.19 when $0.8<|Z|<1.2$\,kpc, 1.63 $\pm$ 0.13 when $1.5<|Z|<2.0$\,kpc,  1.46 $\pm$ 0.12 when $3.0<|Z|<4.0$\,kpc, and  1.59 $\pm$ 0.13 when $5<|Z|<7$\,kpc.
Upon performing a chi-squared test of the hypothesis that the slope is constant as a function of $|Z|$, we find no evidence to reject that hypothesis. 
Consequently, from our analysis, there is no strong evidence that the behavior of the very metal-poor tail of the MDF varies as a function of scale height.

Finally, we note that  metallicities derived from SkyMapper photometry are biased high in the case of carbon-enhanced stars, which would plausibly steepen our MDF since carbon-enhanced stars are more prevalent at low metallicities.
To estimate whether this affect could appreciably alter our MDF, we used the sample of \citet{pfb+14} to investigate the carbon abundance ([C/Fe]) of stars between $-3.0 < $[Fe/H]$ < -2.0$.
We find that the average carbon abundance of stars increases by 0.13\,dex between [Fe/H] = $-2.0$ and [Fe/H] = $-3.0$.
At a typical \teff = 4500\,K, such an increase in the carbon abundance stars with [Fe/H] $= -3.0$ would only increase the photometric metallicity by 0.07\,dex.
Consequently, while we do acknowledge that individual outliers with extremely high carbon abundances would appear to have much higher metallicities, at a population level, the metallicity of a typical star is not significantly affected.

\section{Summary and Conclusion} 
\label{sec:conclusion}

We have presented a low-metallicity map of the Milky Way based on photometric metallicities obtained from new public SkyMapper DR2 data of $\sim$111,000 red giants in the Southern sky with $-3.75\lesssim\mbox{[Fe/H]}\lesssim-0.75$ and $g<17$ out to $\sim7$\,kpc from the solar neighborhood. 
Based on a comparison with new high-resolution spectra as well as using several literature samples and surveys, we demonstrate this sample achieves reliable metallicity precision ($\sigma\sim$0.24\,dex to 0.29\,dex) down to $\mbox{[Fe/H]}\sim-3.3$. 
We display the spatial distribution of stars within the very and extremely metal-poor regimes providing an unprecedented view into the ancient, metal-poor Milky Way. 

We find clear signatures of the components of the Milky Way in our dataset. 
Notably, the thick disk prominently appears in our population of stars with [Fe/H] $> -1.50$ (see Figure~\ref{fig:metallicity_bins}), and the transition from the thick disk to the inner halo is reflected in a sharp drop in metallicities around $|Z| \sim 2.8$\,kpc (see Figure~\ref{fig:metallicity_map}). 
The average metallicities of stars in our sample progressively decrease out to $|Z|\sim7$ kpc. 
At even larger distances, we do not have enough coverage to confidently map out metallicity averages but observe hints of an additional small decrease, as expected for the outer halo. 
It is worth pointing out that a smooth distribution of very and extremely metal-poor stars exists across the entire sky covered by this sample, even within the more metal-rich thick disk. 
This highlights that the most metal-poor stars are well spread across the galaxy, likely owing to their orbital properties that regularly bring to the inner portions of the Galaxy. 
We explore this topic in a separate paper (M. Mardini et al. 2020, in prep), in which we quantify the kinematic properties of our sample of stars and find that $\sim$200 stars with [Fe/H] $< -2.3$ in our sample have orbits consistent with membership to the thick disk.
This result further highlights that very and extremely metal-poor stars exist in the thick disk of the galaxy.

We are also able to quantify the shape of the MDF for $\mbox{[Fe/H]}< -1.5$. 
It is well fit by two exponential profiles with a slope of $\Delta\log(N)/\Delta$[Fe/H] = 0.98$\pm$0.01 for stars with $-2.3<\mbox{[Fe/H]} < -1.5$, and $\Delta\log(N)/\Delta$[Fe/H] = 1.52$\pm$0.05 when $-3.0<\mbox{[Fe/H]}< -2.3$. 
When only including stars within 5.0\,kpc, to account for biases from target selection, we derive slopes of $\Delta\log(N)/\Delta$[Fe/H] = 1.09$\pm$0.02 for stars with $-2.3<\mbox{[Fe/H]} < -1.0$, and $\Delta\log(N)/\Delta$[Fe/H] = 1.53$\pm$0.10 when $-3.0<\mbox{[Fe/H]}< -2.3$. 
In both cases, the MDF steepens at low metallicities, highlighting the difficulty of finding the most metal-poor stars.
More intuitively, the slope from our entire sample implies that the number of metal-poor stars drops by a factor of $\sim$10 for every dex decrease in [Fe/H] when $-2.3<\mbox{[Fe/H]} < -1.0$ and a factor of $\sim$33 for every dex decrease in [Fe/H] when  $-3.0<\mbox{[Fe/H]} < -2.3$.
Unsurprisingly, we also find that the relative frequency of the most metal-poor stars increases as a function of $|Z|$, as 0.8\% of stars in our sample have [Fe/H] $< -2.5$ when $|Z| < 3$\,kpc but comprise 2.7\% of stars when $|Z| > 3$\,kpc. 
Finally, we find some evidence that the MDF for stars with $\mbox{[Fe/H]}< -2.3$ is largely insensitive to scale height $|Z|$ out to $\sim5$\,kpc, suggesting that the most metal-poor stars are spatially distributed in every galactic component. 
Overall, our sample contains $\sim$140 giants with $\mbox{[Fe/H]} < -3.0$, and $\sim$35 with $\mbox{[Fe/H]} < -3.3$.
Detailed follow-up studies of these stars are underway which should help to reveal the full extent of the nature and origin of the ancient component of our Milky Way.


\acknowledgements

A.C. thanks Alex Ji and Kaley Brauer for helpful conversations.
A.C. and A.F. acknowledge support from NSF grant AST-1716251. A.F. thanks the Wissenschaftskolleg zu Berlin for their generous hospitality. 
This work made use of NASA's Astrophysics Data System Bibliographic Services, and the SIMBAD database, operated at CDS, Strasbourg, France. 

This work has made use of data from the European Space Agency (ESA) mission
{\it Gaia} (\url{https://www.cosmos.esa.int/gaia}), processed by the {\it Gaia}
Data Processing and Analysis Consortium (DPAC,
\url{https://www.cosmos.esa.int/web/gaia/dpac/consortium}). Funding for the DPAC
has been provided by national institutions, in particular the institutions
participating in the {\it Gaia} Multilateral Agreement.

The national facility capability for SkyMapper has been funded through ARC LIEF grant LE130100104 from the Australian Research Council, awarded to the University of Sydney, the Australian National University, Swinburne University of Technology, the University of Queensland, the University of Western Australia, the University of Melbourne, Curtin University of Technology, Monash University and the Australian Astronomical Observatory. SkyMapper is owned and operated by The Australian National University's Research School of Astronomy and Astrophysics. The survey data were processed and provided by the SkyMapper Team at ANU. The SkyMapper node of the All-Sky Virtual Observatory (ASVO) is hosted at the National Computational Infrastructure (NCI). Development and support the SkyMapper node of the ASVO has been funded in part by Astronomy Australia Limited (AAL) and the Australian Government through the Commonwealth's Education Investment Fund (EIF) and National Collaborative Research Infrastructure Strategy (NCRIS), particularly the National eResearch Collaboration Tools and Resources (NeCTAR) and the Australian National Data Service Projects (ANDS).

 \newcommand{\noop}[1]{}

\end{document}